\documentclass[twocolumn,showpacs,preprintnumbers,amsmath,amssymb]{revtex4}

\usepackage{graphics}
\usepackage{amsmath}
\usepackage{amstext}
\usepackage{latexsym}
\usepackage{epsfig}
\usepackage{amsfonts}
\usepackage{amssymb}
\usepackage{amscd}
\usepackage{mathcomp}

\def\z{`}
\def\vv{``}
\def\t{~}

\begin{document}

\title{Eavesdropping on Blind Quantum Key Distribution\\
through a Labeling Attack}

\author{Marco Lucamarini}
\affiliation{Dipartimento di Fisica, Universit\`a di Camerino,
I-62032 Camerino, Italy}


\begin{abstract}
I present an eavesdropping on the protocol proposed by W.-H. Kye,
\textit{et al.} [Phys. Rev. Lett. \textbf{95}, 040501 (2005)]. I
show how an undetectable Eve can steal the whole information by
labeling and then measuring the photons prepared by the user
Alice.
\end{abstract}

\pacs{03.67.Dd, 03.67.Hk}

\maketitle

In Ref.\t\cite{Kye05} it is described a novel three-way protocol
for quantum key distribution, defined \vv blind" because it does
not require the public revelation of the polarization bases; let
me call it for ease KKKP, after the names of its inventors. One
nice feature of KKKP is that the \vv control mode" present in
Ping-Pong-like protocols is not necessary, because of the
randomness of the traveling photons polarization. Unfortunately
this same feature makes KKKP vulnerable to a number of attacks by
an eavesdropper (Eve) whose aim is to steal information from Alice
and Bob. As a consequence KKKP experienced a series of
modifications\t\cite{Kye05,Kye05a,Kye05b} in order to improve its
security without changing the philosophy of the scheme: the
Impersonation attack\t\cite{Dusek99} is avoided through the
preparation of two pulses on Alice side followed, at a proper
time, by a random blocking of one of the two, and through the
addition of a random parameter on Bob side, the \z shuffling
factor' $s$\t\cite{Kye05}. Notwithstanding, these changes alone
are not sufficient to avert the attack, as noticed
in\t\cite{Zhang05}, and it was necessary to double the shuffling
parameter in two, the shufflings $s_1$ and $s_2$\t\cite{Kye05b}.
Finally, an eavesdropping resulting from the combination of
Invisible-photon attack\t\cite{Cai05} and the PNS
attack\t\cite{Gisin02} can be avoided through a random check of
the intensities of the beams by Alice and/or
Bob\t\cite{Kye05,Kye05a}.

In this Comment I show that despite the above modifications KKKP
can still be attacked successfully without risk of detection. The
main tool at Eve's disposal is what I call a \vv
labeling-and-measure" strategy. It can be thought as a small shift
of one of the involved photons' degree of freedom, like the
wavelength for instance, in a way similar to what proposed by Cai
in\t\cite{Cai05}. This makes the photons distinguishable to Eve,
who can thus measure the relevant parameters of the communication.
\\
\indent In the three-way KKKP Alice prepares a photon in a
polarization state $\left\vert \psi_{1}\right\rangle
=\cos\theta\left\vert 0\right\rangle -\sin\theta\left\vert
1\right\rangle $, choosing randomly the angle $\theta$. Bob
rotates the polarization of the received photon by applying the
unitary operator $\widehat{U}_{y}
(\phi)=\cos\phi\widehat{I}-i\sin\phi\widehat{\sigma}_{y}$, with
$\phi$ randomly chosen as well; then returns the photon to Alice.
Alice applies the transformation $\widehat{U}_{y}(-\theta+\left(
-1\right)  ^{k}\pi/4)$ to the photon, where $k=0,1$, thus removing
the protection $\theta$ from the state, and encoding on it the key
\z $k$'. In this way Bob, after compensating for his angle $\phi$,
can deterministically infer the key. In the stronger version of
KKKP Alice sends two photons to Bob, with independent angles
$\theta_{1}$ and $\theta_{2}$. Bob executes $\widehat
{U}_{y}(\phi+\left( -1\right) ^{s_1}\pi/4)$ on the first and
$\widehat {U}_{y}(\phi+\left( -1\right) ^{s_2}\pi/4)$ on the
second, and returns them to Alice. Alice adds the key as before,
but she also blocks one of the two pulses according to the value
$0$ or $1$ of the \z blocking factor' $b$. Then, after removing
any temporal label from the resulting photon, she forwards it to
Bob, who measures it as before. Afterward, Alice's public
disclosure of $b$ allows Bob to reconstruct the key. To this
stronger version of KKKP I add also the random control on the
intensity of the beams during the protocol, by Alice and/or Bob,
and call this final version \underline{KKKP}.
\smallskip

Now I describe the label-and-measure eavesdropping on
\underline{KKKP}. For explicative purposes the label I use is the
wavelength of the photons. However any other degree of freedom
different from the polarization (e.g. the momentum direction, an
eventual phase respect to a reference signal, the temporal
distribution, etc.) can be used as a label by Eve, and
constitutes a loophole of the protocol.\\
\indent Let us suppose, like in the Impersonation attack, that
there are Eve1, who impersonates Bob to Alice, and Eve2, who
impersonates Alice to Bob. The hard work is made by Eve1, who
steals both the key $k$ and the block factor $b$ from Alice. In
\underline{KKKP} Alice sends out the pulses $\theta_{1}$ and
$\theta_{2}$ at a certain wavelength $\lambda$. Eve1 chooses
randomly one of the two photons, say the first one ($p1$), and
shifts its wavelength into $\lambda_{1} =\lambda+\delta$, while
leaving the second ($p2$) unchanged. $\delta$ is a quantity
different from zero whose smallness depends on Eve1's technology,
which one must assume to be infinite, or at least greater than
Alice and Bob's one. According to the protocol, Alice performs the
operations $\widehat{U}_{y}(-\theta_{1}+\left( -1\right)
^{k}\pi/4)$ on $p1$ and $\widehat{U}_{y}(-\theta_{2}+\left(
-1\right) ^{k}\pi/4)$ on $p2$. Then she blocks one of the two
photons, say $p2$, and forwards $p1$, after delaying it properly
to make it undistinguishable in time from $p2$. It is
straightforward to see that from Alice side exits only $p1$ in the
state:
\begin{equation}\label{State}
    \left\vert \xi\right\rangle
=\left\vert \left( -1\right) ^{k}\pi/4,\lambda_1\right\rangle
_{p1},
\end{equation}
where the wavelength is indicated together with the polarization
state. Note that the protection $\theta_{1}$ has been removed from
$p1$ by Alice through her last transformations. By measuring the
wavelength and the polarization of the photon in the
state\t\eqref{State} Eve1 can infer which pulse was blocked, i.e.
the blocking factor $b$, and the value of $k$. These two values
are accordingly used by Eve2 on a pair of fake pulses sent to Bob,
in order to mimic to him the presence of Alice. Since the
imitation is perfect Bob has no chance to unveil the attack. Let
me point out that although I described the eavesdropping as two
separate temporal sequences it is on the contrary important that
Eve1 and Eve2 act in synchronization in the following way:

\smallskip

\noindent \underline{1}. Eve1 modifies $p1$ and stores $p1$ and
$p2$\\
\noindent \underline{2}. Eve2 sends a pair of fake pulses to Bob\\
\noindent \underline{3}. when Eve2 receives back the two fake
pulses from Bob\\
\indent Eve1 forwards to Alice the two photons she stored\\
\noindent \underline{4}. ...and so on.

\smallskip

\noindent This makes the usage of public receipts of the photons
by Alice and Bob ineffective against Eve. Finally, a control of
the intensity of the beams is clearly useless in this case since
the intensity is not altered by Eve.\smallskip

In conclusion I showed an explicit label-and-measure eavesdropping
against the stronger version of the KKKP protocol\t\cite{Kye05},
illustrated by a simple wavelength shift of one of the photons
prepared by Alice. This attack provides Eve with a complete
knowledge of the exchanged information, without the possibility by
Alice and Bob of detect her presence. The insertion of
(wavelength) filters in analogy with what proposed
in\t\cite{Cai05} does not work in this case because there is no
polarization detections during the protocol. Nevertheless, the
introduction of polarization controls along the channels of KKKP
is not simple, just because of the randomness of the polarization
bases.\\
\indent The solution of the problem should include the Alice
removal of all potential Eve's labels. Perhaps Alice could use the
teleportation process to transfer the polarization state of the
unblocked photon on another photon whose degrees of freedom are
not controlled by Eve. Despite its practical difficulty this
solution seems to be not outside the range of current technology.

\bibliography{Blind}

\end{document}